\documentclass[12pt]{article}
\usepackage[margin=1in]{geometry}
\usepackage{amsmath}
\usepackage{amssymb}
\usepackage{graphicx}
\usepackage{float}
\usepackage{tikz}

\newcommand{\MySpace}{2.5}

\title{Proposal for a Correction to the Temporal Correlation Coefficient Calculation for Temporal Networks}
\author{Pigott F., Herrera M.}

\begin{document}
\maketitle

\begin{abstract}
Measuring the topological overlap of two graphs becomes important when assessing the changes between temporally adjacent graphs in a time-evolving network. Current methods depend on the fraction of nodes that have persisting edges. This breaks down when there are nodes with no edges, persisting or otherwise. The following outlines a proposed correction to ensure that correlation metrics have the expected behavior.
\end{abstract}

\section{Previous Proposal}
\cite{1} defines the \emph{topological overlap} in the neighborhood of \(i\) between two consecutive time steps \([t_m,t_{m+1}]\) as:
\begin{equation}
C_i(t_m,t_{m+1}) = \frac{ \sum_j a_{ij}(t_m)a_{ij}(t_{m+1})}{\sqrt{[ \sum_j a_{ij}(t_m)][ \sum_j a_{ij}(t_{m+1})]}}
\label{eq:Ci}
\end{equation}

where \(a_{ij}\) represents an entry in the unweighted adjacency matrix of the graph, so summing over \(a_{ij}\) gives the interactions between \(i\) and every other node.

Then the \emph{average topological overlap} of the neighborhood of node \(i\) as the average of \(C_i(t_m,t_{m+1})\) over all possible subsequent temporal snapshots:

\begin{equation}
C_i = \frac{1}{M-1} \sum_{m=1}^{M-1} C_i(t_m,t_{m+1})
\label{eq:Ci1}
\end{equation} 

The average topological overlap of all \(C_i\) can be said to represent the \emph{temporal clustering} of all of the edges in the network, which we will call the \emph{temporal clustering coefficient} (TCC).

\begin{equation}
C = \frac{1}{N}\sum_{i = 1}^{N} C_i
\label{eq:C1}
\end{equation}

Identically, the order of summations in this formulation can be reversed (simply to make the following more obvious), to give the \emph{average topological overlap} of the graph at \(t_m\) with the subsequent graph at \(t_{m+1}\):
\begin{equation}
C_m = \frac{1}{N} \sum_{i = 1}^{N} C_i(t_m,t_{m+1})
\label{eq:Cm1}
\end{equation}

and the same C is then given by:
\begin{equation}
C = \frac{1}{M-1}\sum_{m=1}^{M-1} C_m
\label{eq:C1b}
\end{equation}

According to \cite{1} this formulation gives \(C_m=1\) if and only if the graphs at \(t_m\) and \(t_{m+1}\)  have exactly the same configuration of edges, and \(C_m = 0\) if the graph at \(t_m\) and \(t_{m+1}\) do not share any edges.  This claim is only true if all of the \(N\) nodes considered in the calculation have at least one edge. 

Equation (\ref{eq:Ci}) results in an undefined \(\frac{0}{0}\) in the case where the node \(i\) has no edges in one or both time steps \(t_m\) or \(t_{m+1}\). If that undefined value is simply set to zero (choose zero because the node \(i_m\) shares no edges with the node \(i_{m+1}\)), \(C\) between identical unconnected graphs is equal to the fraction of connected nodes in the graph. The formula by \cite{1} provides no method for dealing with networks where \(N(t_m)\) (the number of nodes participating in the network) changes in time.

This presents a significant problem, because for small time snapshots, many temporal graphs are often unconnected (contain unconnected nodes) \cite{1} and \(N(t_m)\) changes constantly with time. For unconnected graphs, \(C_m\) describing the relationship between two identical graphs gives \(C_m \neq 1\), and can significantly underestimate the correlation between two graphs.

\section{Proposed Correction}
To ensure that \(C_m\) has the expected behavior when the graph is unconnected, a constant \(n\) is replaced by the maximum number of connected nodes in the network for the two time steps being compared. Using \(\max [N(t_m),N(t_{m+1})]\) rather than simply \(N(t_m)\) or \(N(t_{m+1}\) ensures that \(C < 1\) for all non-identical graphs, using \(\min [N(t_m),N(t_{m+1})]\) would give \(C = 1\) for disconnected graphs where the only change is an edge appearing or disappearing. Be aware that this method will still give a \(\frac{0}{0}\) for correlation between two graphs both with zero edges.

\begin{equation}
C_m = \frac{1}{\max [N(t_m),N(t_{m+1})]} \sum_{i = 1}^{N} C_i(t_m,t_{m+1})
\label{eq:Cm2}
\end{equation}

\begin{equation}
C = \frac{1}{M-1}\sum_{m=1}^{M-1}  \left( \frac{1}{\max [N(t_m),N(t_{m+1})]} \sum_{i = 1}^{N} C_i(t_m,t_{m+1}) \right)
\label{eq:C2}
\end{equation}

\newpage
\section{Examples}

Below are some simple examples to show the difference between the two methods. 

\subsection{Connected graph}
\begin{figure}[H]
\centering

\begin{tikzpicture}[scale=2.5]

\draw (0,0) -- (.5,.84);
\draw (1,0) -- (0,0);
\draw [green, fill] (.5,.84) circle [radius=0.1];
\node at (.5,.84) {1};
\draw [green, fill] (0,0) circle [radius=0.1];
\node at (0,0) {2};
\draw [green, fill] (1,0) circle [radius=0.1];
\node at (1,0) {3};

\node at (.5,-.25) {\(t_m\)};

\draw (\MySpace + 0,0) -- (\MySpace + .5,.84);
\draw (\MySpace + .5,.84) -- (\MySpace + 1,0);
\draw [green, fill] (\MySpace + .5,.84) circle [radius=0.1];
\node at (\MySpace + .5,.84) {1};
\draw [green, fill] (\MySpace + 0,0) circle [radius=0.1];
\node at (\MySpace + 0,0) {2};
\draw [green, fill] (\MySpace + 1,0) circle [radius=0.1];
\node at (\MySpace + 1,0) {3};

\node at (\MySpace + .5,-.25) {\(t_{m+1}\)};
\end{tikzpicture}
\caption{A connected graph. Here Methods 1 and 2 give the same result.}
\end{figure}
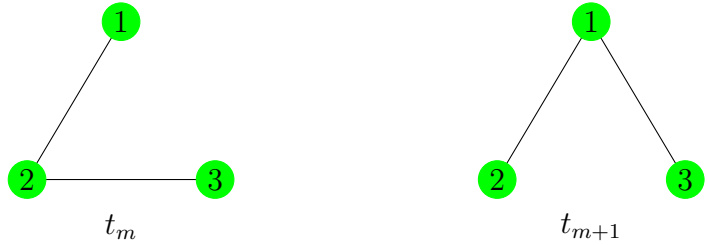

%
 \emph{topological overlap} in the neighborhood of \(i\) between two consecutive time steps:
\[\text{(\ref{eq:Ci}): } C_{i} (t_m,t_{m+1}): C_{i=1} (t_m,t_{m+1})  = \frac{1}{\sqrt{2}}, \; C_{i=2} (t_m,t_{m+1}) = \frac{1}{\sqrt{2}}, \; C_{i=3} (t_m,t_{m+1}) =  0\]

 \begin{flushleft}
 { \bfseries Method 1: }\( C = \frac{1}{N} \sum_{i=1}^N \left( \frac{1}{M-1} \sum_{m=1}^{M-1} C_i(t_m,t_{m+1}) \right) = \frac{1}{3} \left( \frac{1}{1} (\frac{1}{\sqrt{2}}+\frac{1}{\sqrt{2}}+0) \right)  = \frac{\sqrt{2}}{3} \)
 
{ \bfseries Method 2:} \(C = \frac{1}{M-1}\sum_{m=1}^{M-1}  \left( \frac{1}{\max [N(t_m),N(t_{m+1})]} \sum_{i = 1}^{N} C_i(t_m,t_{m+1}) \right) = \frac{1}{1} \left( \frac{1}{3} (\frac{1}{\sqrt{2}}+\frac{1}{\sqrt{2}}+0)\right) =  \frac{\sqrt{2}}{3} \)
 \end{flushleft}

\subsection{Unconnected graph}
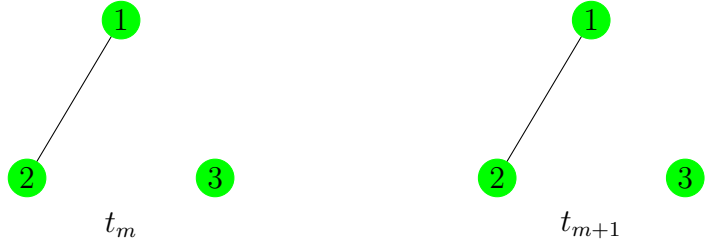
\begin{figure}[H]
\centering
\begin{tikzpicture}[scale=2.5]

\draw (0,0) -- (.5,.84);
\draw [green, fill] (.5,.84) circle [radius=0.1];
\node at (.5,.84) {1};
\draw [green, fill] (0,0) circle [radius=0.1];
\node at (0,0) {2};
\draw [green, fill] (1,0) circle [radius=0.1];
\node at (1,0) {3};

\node at (.5,-.25) {\(t_m\)};

\draw (\MySpace + 0,0) -- (\MySpace + .5,.84);
\draw [green, fill] (\MySpace + .5,.84) circle [radius=0.1];
\node at (\MySpace + .5,.84) {1};
\draw [green, fill] (\MySpace + 0,0) circle [radius=0.1];
\node at (\MySpace + 0,0) {2};
\draw [green, fill] (\MySpace + 1,0) circle [radius=0.1];
\node at (\MySpace + 1,0) {3};

\node at (\MySpace + .5,-.25) {\(t_{m+1}\)};

\end{tikzpicture}
\caption{An unconnected graph. As the graphs are identical the expected value is one, but method 1 for calculating \(C\) gives \(\frac{2}{3}\), or the fraction of nodes that are participating in the network at time \(t_m\).}
\end{figure}


\emph{topological overlap} in the neighborhood of \(i\) between two consecutive time steps:
\[\text{(\ref{eq:Ci}): } C_{i} (t_m,t_{m+1}): C_{i=1} (t_m,t_{m+1})  = 1, \; C_{i=2} (t_m,t_{m+1}) = 1, \; C_{i=3} (t_m,t_{m+1}) = \frac{0}{0} = 0\]
 
 \begin{flushleft}
 { \bfseries Method 1: }\( C = \frac{1}{N} \sum_{i=1}^N \left( \frac{1}{M-1} \sum_{m=1}^{M-1} C_i(t_m,t_{m+1}) \right) = \frac{1}{3}(\frac{1}{1}(1+1+0)) = \frac{2}{3} \)
 
{ \bfseries Method 2:} \(C = \frac{1}{M-1}\sum_{m=1}^{M-1}  \left( \frac{1}{\max [N(t_m),N(t_{m+1})]} \sum_{i = 1}^{N} C_i(t_m,t_{m+1}) \right) = \frac{1}{1} \left( \frac{1}{2} (1 + 1+ 0) \right) =  1\)
 \end{flushleft}

\newpage 
\subsection{Time series with unconnected graphs}

\begin{figure}[H]
\centering
\begin{tikzpicture}[scale=2.5]

\draw (0,0) -- (0,1);
\draw (0,1) -- (1,0);
\draw (0,1) -- (1,1);
\draw [green, fill] (0,1) circle [radius=0.1];
\node at (0,1) {1};
\draw [green, fill] (0,0) circle [radius=0.1];
\node at (0,0) {2};
\draw [green, fill] (1,0) circle [radius=0.1];
\node at (1,0) {3};
\draw [green, fill] (1,1) circle [radius=0.1];
\node at (1,1) {4};

\node at (.5,-.25) {\(t_m\)};

\draw (\MySpace + 0,0) -- (\MySpace + 0,1);
\draw (\MySpace + 0,1) -- (\MySpace + 1,0);
\draw [green, fill] (\MySpace + 0,1) circle [radius=0.1];
\node at (\MySpace + 0,1) {1};
\draw [green, fill] (\MySpace + 0,0) circle [radius=0.1];
\node at (\MySpace + 0,0) {2};
\draw [green, fill] (\MySpace + 1,0) circle [radius=0.1];
\node at (\MySpace + 1,0) {3};
\draw [green, fill] (\MySpace + 1,1) circle [radius=0.1];
\node at (\MySpace + 1,1) {4};

\node at (\MySpace + .5,-.25) {\(t_{m+1}\)};

\draw (2*\MySpace + 0,0) -- (2*\MySpace + 0,1);
\draw [green, fill] (2*\MySpace + 0,1) circle [radius=0.1];
\node at (2*\MySpace + 0,1) {1};
\draw [green, fill] (2*\MySpace + 0,0) circle [radius=0.1];
\node at (2*\MySpace + 0,0) {2};
\draw [green, fill] (2*\MySpace + 1,0) circle [radius=0.1];
\node at (2*\MySpace + 1,0) {3};
\draw [green, fill] (2*\MySpace + 1,1) circle [radius=0.1];
\node at (2*\MySpace + 1,1) {4};

\node at (2*\MySpace + .5,-.25) {\(t_{m+2}\)};

\end{tikzpicture}
\caption{A network containing four nodes (\(N = 4\)) which becomes disconnected during the observed time window (\(N(t_m) = 4\), however, \(N(t_{m+1}) = 3\) and \(N(t_{m+2}) = 2\)). Note three time steps, so \(M = 3\).}
\end{figure}
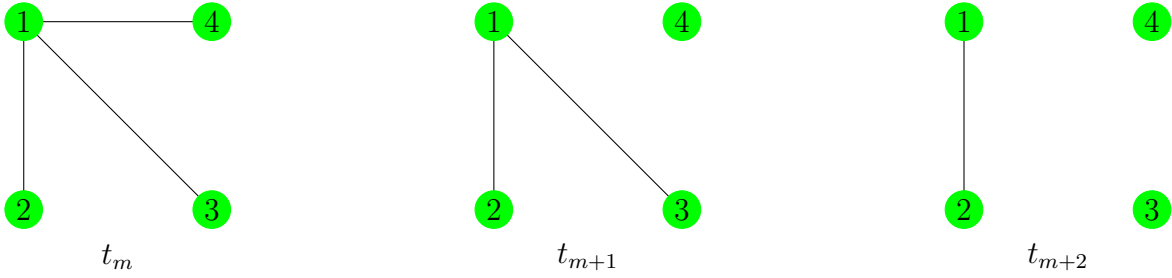

\begin{table}[H]
\begin{tabular}{l l l l}
  (\ref{eq:Ci}): \(C_{i} (t_m,t_{m+1})\): & \(C_{i} (t_{m+1},t_{m+2})\):   \\ \hline
  \(C_{i=1} (t_m,t_{m+1})  = \frac{2}{\sqrt{6}}\) &   \(C_{i=1} (t_{m+1},t_{m+2})  =\frac{1}{\sqrt{2}}\) \\ 
    \(C_{i=2} (t_m,t_{m+1}) = 1  \)  &   \(C_{i=2} (t_{m+1},t_{m+2}) = 1  \) \\  
  \(C_{i=3} (t_m,t_{m+1}) = 1 \) &   \(C_{i=3} (t_{m+1},t_{m+2}) = \frac{0}{0} = 0  \)  \\ 
  \(C_{i=4} (t_m,t_{m+1}) = \frac{0}{0} = 0 \) &   \(C_{i=4} (t_{m+1},t_{m+2}) = \frac{0}{0} = 0  \) 
 \end{tabular}
 \caption{\emph{topological overlap} in the neighborhood of \(i\) between two consecutive time steps}
 \end{table}


\begin{table}[H]
\begin{tabular}{ l r }
\multicolumn{2}{ l }{\bf{Method 1:}} \\ \hline
(\ref{eq:Cm1}):  \(C_m = \frac{1}{N} \sum_{i = 1}^{N} C_i(t_m,t_{m+1}) \) & (\ref{eq:C1b}): \( C = \frac{1}{M-1} \sum_m^{M-1} C_m\)\\ [1.2ex]
\(C_{m} = \frac{1}{4}\sum_{i=1}^4 C_i(t_m,t_{m+1}) \approx 0.70  \) & \( \frac{1}{4}\sum_{m}^{3-1} C_m\)\\ [1.2 ex]
\(C_{m+1} =   \frac{1}{4}\sum_{i=1}^4 C_i(t_{m+1},t_{m+2}) \approx 0.45 \) &\( \approx 0.57 \)  \\  
\end{tabular}
\end{table}
 
\begin{table}[H]
\begin{tabular}{ l r }
\multicolumn{2}{ l }{\bf{Method 2:}} \\ \hline
(\ref{eq:Cm2}):  \(C_m = \frac{1}{\max [N(t_m),N(t_{m+1})]} \sum_{i = 1}^{N} C_i(t_m,t_{m+1}) \) & (\ref{eq:C2}): \( C = \frac{1}{M-1} \sum_m^{M-1} C_m\)\\ [1.2ex]
\(C_{m} = \frac{1}{4}\sum_{i=1}^4 C_i (t_m,t_{m+1}) \approx 0.70  \) & \( \frac{1}{3-1}\sum_{m}^{3-1} C_m\)\\ [1.2 ex]
\(C_{m+1} =   \frac{1}{3}\sum_{i=1}^4 C_i (t_{m+1},t_{m+2}) \approx 0.57 \) &\( \approx 0.64 \)  \\  
\end{tabular}
\end{table}

\subsection{Time series with identical unconnected graphs}

\begin{figure}[H]
\centering
\begin{tikzpicture}[scale=2.5]

\draw (0,0) -- (0,1);
\draw (0,1) -- (1,0);
\draw (0,1) -- (1,1);
\draw [green, fill] (0,1) circle [radius=0.1];
\node at (0,1) {1};
\draw [green, fill] (0,0) circle [radius=0.1];
\node at (0,0) {2};
\draw [green, fill] (1,0) circle [radius=0.1];
\node at (1,0) {3};
\draw [green, fill] (1,1) circle [radius=0.1];
\node at (1,1) {4};

\node at (.5,-.25) {\(t_m\)};

\draw (\MySpace + 0,0) -- (\MySpace + 0,1);
\draw (\MySpace + 0,1) -- (\MySpace + 1,0);
\draw [green, fill] (\MySpace + 0,1) circle [radius=0.1];
\node at (\MySpace + 0,1) {1};
\draw [green, fill] (\MySpace + 0,0) circle [radius=0.1];
\node at (\MySpace + 0,0) {2};
\draw [green, fill] (\MySpace + 1,0) circle [radius=0.1];
\node at (\MySpace + 1,0) {3};
\draw [green, fill] (\MySpace + 1,1) circle [radius=0.1];
\node at (\MySpace + 1,1) {4};

\node at (\MySpace + .5,-.25) {\(t_{m+1}\)};

\draw (2*\MySpace + 0,0) -- (2*\MySpace + 0,1);
\draw (2*\MySpace + 0,1) -- (2*\MySpace + 1,0);
\draw [green, fill] (2*\MySpace + 0,1) circle [radius=0.1];
\node at (2*\MySpace + 0,1) {1};
\draw [green, fill] (2*\MySpace + 0,0) circle [radius=0.1];
\node at (2*\MySpace + 0,0) {2};
\draw [green, fill] (2*\MySpace + 1,0) circle [radius=0.1];
\node at (2*\MySpace + 1,0) {3};
\draw [green, fill] (2*\MySpace + 1,1) circle [radius=0.1];
\node at (2*\MySpace + 1,1) {4};

\node at (2*\MySpace + .5,-.25) {\(t_{m+2}\)};

\end{tikzpicture}
\caption{A graph which becomes unconnected during the observed time window. Note that even though the second two time steps are identical, the \(C\) between \(t_m\) and \(t_{m+1}\) \(\neq 1\) because the graph is disconnected. Because of this, \(C\) for the time series of graphs is underestimated when performing the calculation using Method 1.}
\label{fig:4}
\end{figure}
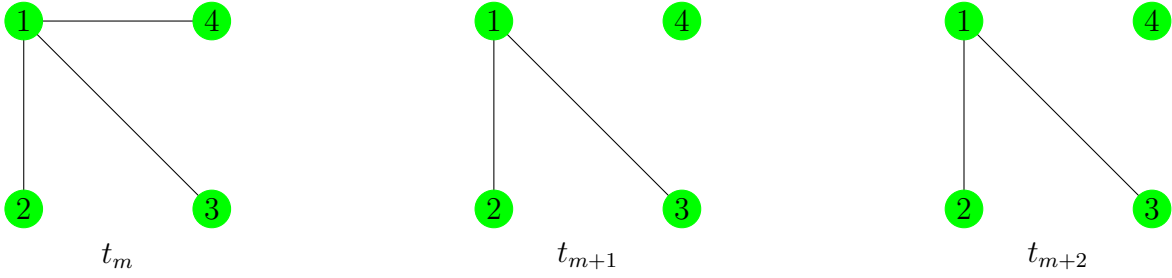

\begin{table}[H]
\begin{tabular}{l l l l}
 (\ref{eq:Ci}): \(C_{i} (t_m,t_{m+1})\): & \(C_{i} (t_{m+1},t_{m+2})\):   \\ \hline
  \(C_{i=1} (t_m,t_{m+1})  = \frac{2}{\sqrt{6}}\) &   \(C_{i=1} (t_{m+1},t_{m+2})  =1\) \\ 
    \(C_{i=2} (t_m,t_{m+1}) = 1  \)  &   \(C_{i=2} (t_{m+1},t_{m+2}) = 1  \) \\  
  \(C_{i=3} (t_m,t_{m+1}) = 1 \) &   \(C_{i=3} (t_{m+1},t_{m+2}) = 1  \)  \\ 
  \(C_{i=4} (t_m,t_{m+1}) = \frac{0}{0} = 0 \) &   \(C_{i=4} (t_{m+1},t_{m+2}) = \frac{0}{0} = 0  \) 
 \end{tabular}
 \caption{\emph{topological overlap} in the neighborhood of \(i\) between two consecutive time steps}
 \end{table}
 
 
 \begin{table}[H]
\begin{tabular}{ l r }
\multicolumn{2}{ l }{\bf{Method 1:}} \\ \hline
(\ref{eq:Cm1}):  \(C_m = \frac{1}{N} \sum_{i = 1}^{N} C_i(t_m,t_{m+1}) \) & (\ref{eq:C1b}): \( C = \frac{1}{M-1} \sum_m^{M-1} C_m\)\\ [1.2ex]
\(C_{m} = \frac{1}{4}\sum_{i=1}^4 C_i (t_{m},t_{m+1})  \approx  .70 \) & \( \frac{1}{4}\sum_{m}^{3-1} C_m\)\\ [1.2 ex]
\(C_{m+1} =   \frac{1}{4}\sum_{i=1}^4 C_i(t_{m+1},t_{m+2})  =  \frac{3}{4} \) &\( \approx 0.73 \)  \\  
\end{tabular}
\caption{As graphs \(t_{m+1}\) and \(t_{m+2}\) are identical, \(C_{m+1}\) should be equal to one, but using Method 1 the calculation gives \(\frac{3}{4}\), the fraction of nodes participating in the network at time \(t_{m+1}\).}
\end{table}
 
\begin{table}[H]
\begin{tabular}{ l r }
\multicolumn{2}{ l }{\bf{Method 2:}} \\ \hline
(\ref{eq:Cm2}):  \(C_m = \frac{1}{\max [N(t_m),N(t_{m+1})]} \sum_{i = 1}^{N} C_i(t_m,t_{m+1}) \) & (\ref{eq:C2}): \( C = \frac{1}{M-1} \sum_m^{M-1} C_m\)\\ [1.2ex]
\(C_{m} = \frac{1}{4}\sum_{i=1}^4 C_i (t_{m},t_{m+1})  \approx .70  \) & \( \frac{1}{3-1}\sum_{m}^{3-1} C_m\)\\ [1.2 ex]
\(C_{m+1} =   \frac{1}{3}\sum_{i=1}^4 C_i(t_{m+1},t_{m+2})  = 1 \) &\( \approx 0.85 \)  \\  
\end{tabular}
\end{table}

If this time series of graphs were extended, with \(t_{m+2}, t_{m+3}...\) identical to \(t_{m+1}\), the time series would logically demonstrate a very high (asymptotically 1) temporal correlation, as the graphs hardly ever change. Calculated using the Method 2 formulation, \(C \rightarrow 1\), but using the Method 1 formulation, \(C\) is asymptotically equal to the average fraction of nodes participating in the network.

\begin{figure}[H]
\includegraphics[width = .8\textwidth]{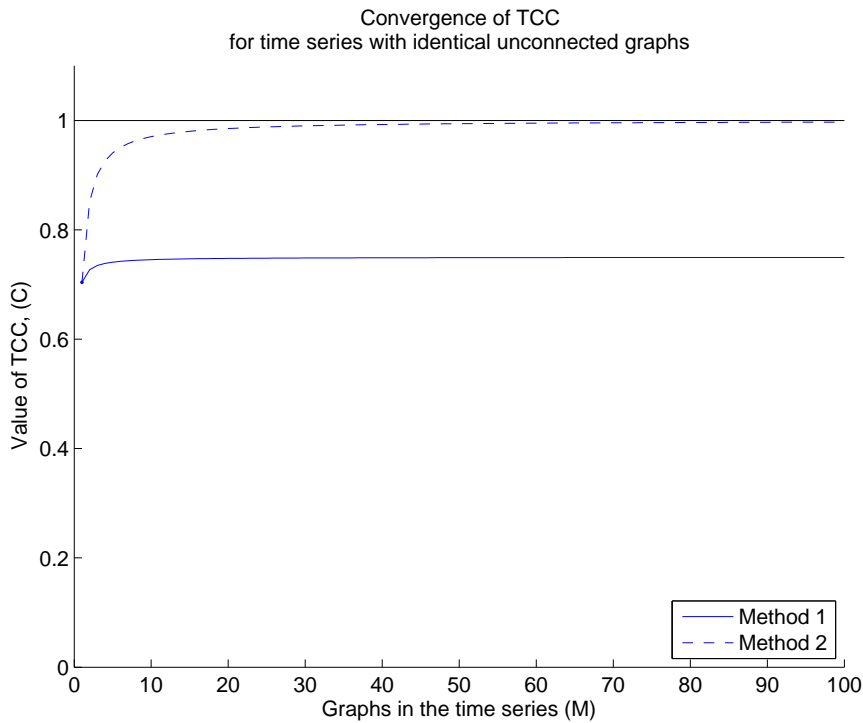}
\caption{Plot showing convergence behavior of TCC for Methods 1 and 2 for the time series described above, adding identical unconnected graphs to the series in figure (\ref{fig:4}).}
\end{figure}

\section{Conclusion}
Because Method 1 for calculating the temporal correlation coefficient relies on a fixed number of nodes in the network \(N\), some modifications (presented as Method 2) need to be made to the formulation to avoid systematically underestimating the correlation between two unconnected graphs. In a time series of graphs which have, on average, unconnected nodes, the temporal correlation coefficient \(C\) calculated using Method 1 will underestimate the correlation between two graphs by the fraction \(\frac{\bar{N_0}}{N}\) where \(\bar{N_0}\) is the average number of unconnected nodes over time and \(N\) is the total number of nodes considered.

\end{document}